# Toxic Code Snippets on Stack Overflow

Chaiyong Ragkhitwetsagul, Jens Krinke, Matheus Paixao, Giuseppe Bianco, Rocco Oliveto

**Abstract**—Online code clones are code fragments that are copied from software projects or online sources to Stack Overflow as examples. Due to an absence of a checking mechanism after the code has been copied to Stack Overflow, they can become toxic code snippets, e.g., they suffer from being outdated or violating the original software license. We present a study of online code clones on Stack Overflow and their toxicity by incorporating two developer surveys and a large-scale code clone detection. A survey of 201 high-reputation Stack Overflow answerers (33% response rate) showed that 131 participants (65%) have ever been notified of outdated code and 26 of them (20%) rarely or never fix the code. 138 answerers (69%) never check for licensing conflicts between their copied code snippets and Stack Overflow's CC BY-SA 3.0. A survey of 87 Stack Overflow visitors shows that they experienced several issues from Stack Overflow answers: mismatched solutions, outdated solutions, incorrect solutions, and buggy code. 85% of them are not aware of CC BY-SA 3.0 license enforced by Stack Overflow, and 66% never check for license conflicts when reusing code snippets. Our clone detection found online clone pairs between 72,365 Java code snippets on Stack Overflow and 111 open source projects in the curated Qualitas corpus. We analysed 2,289 non-trivial online clone candidates. Our investigation revealed strong evidence that 153 clones have been copied from a Qualitas project to Stack Overflow. We found 100 of them (66%) to be outdated, of which 10 were buggy and harmful for reuse. Furthermore, we found 214 code snippets that could potentially violate the license of their original software and appear 7,112 times in 2,427 GitHub projects.

✦

## 1 INTRODUCTION

Stack Overflow is a popular online programming community with 7.6 million users, 14 million questions, and 23 million answers[1]. It allows programmers to ask questions and give answers to programming problems. The website has found to be useful for software development [16], [33], [49], [52], [53], [69], [70], [74] and also valuable for educational purposes [47]. On Stack Overflow, each conversation contains a question and a list of answers. The answers frequently contain at least one code snippet as a solution to the question asked. We found that the code snippets are often not authored directly on the Stack Overflow website but copied from another location. A snippet in an answer could be copied and modified from a code snippet in the question, copied from the answerer's own code or from other locations including open source software (OSS) systems.

The process of posting and answering questions on Stack Overflow that involves the reuse (copying) of source code can be considered code cloning. Code cloning is an activity of reusing source code by copying and pasting. It normally occurs in software development and account from 7% to 23% of source code in typical software systems [8]. The benefits and drawbacks of clones are still controversial. Several authors state that clones lead to bug propagations and software maintenance issues [29], while some others suggest that clones are not harmful and can even be beneficial [30], [63].

Code cloning can also have side effects such as violating software licenses or introducing software vulnerabilities. Carelessly cloning code from one project to another project with a different license may cause a software license violation [22]. This also happens within the context of online Q&A websites such as Stack Overflow. An et al. [3] showed that 1,279 cloned snippets between Android apps and Stack Overflow potentially violate software licenses. Security is also among the main concerns when the code is copied from an online source. For example, Stack Overflow helps developers to solve Android programming problems more quickly than other resources while, at the same time, offers less secure code than books or the official Android documentation [2].

We call code snippets that are copied from software systems to online Q&A websites (such as Stack Overflow) and vice versa as "online code clones." There are two directions in creating online code clones: (1) code is cloned from a software project to a Q&A website as an example; or (2) code is cloned from a Q&A website to a software project to obtain a functionality, perform a particular task, or fixing a bug. Similar to classic code clones, i.e., clones between software systems, online code clones can lead to license violations, bug propagation, an introduction of vulnerabilities, and re-use of outdated code. Unfortunately, online clones are difficult to locate and fix since the search space in online code corpora is larger and no longer confined to a local repository.

To have a deeper insight into online code clones, we surveyed 201 high-reputation Stack Overflow answerers. The results of such a survey show that online code cloning occurs on Stack Overflow. Stack Overflow answerers frequently clone code from other locations, such as their personal projects, company projects, and open source projects, to Stack Overflow as a solution or a complement to a solution. The code cloning activity on Stack Overflow is obviously beneficial considered the popularity of Stack Overflow and its influence on software development [49], [52], [53]. On the other hand, there is also a downside caused by low quality, defective, and harmful code snippets that are reused without awareness by millions of users [2], [20], [83].

---
1. Data as of 21 August 2017 from https://stackexchange.com/sites

```
/* Code in Stack Overflow post ID 22315734 */
public int compare(byte[] b1,int s1,int l1, ...) {
  try {
    buffer.reset(b1,s1,l1); /* parse key1 */
    key1.readFields(buffer);
    buffer.reset(b2,s2,l2); /* parse key2 */
    key2.readFields(buffer);
  } catch (IOException e) {
    throw new RuntimeException(e);
  }
  return compare(key1,key2); /* compare them */
}
```

```
/* WritableComparator.java (2014-11-21) */
public int compare(byte[] b1,int s1,int l1, ...) {
  try {
    buffer.reset(b1,s1,l1); /* parse key1 */
    key1.readFields(buffer);
    buffer.reset(b2,s2,l2); /* parse key2 */
    key2.readFields(buffer);
    buffer.reset(null,0,0); /* clean up reference */
  } catch (IOException e) {
    throw new RuntimeException(e);
  }
  return compare(key1, key2); /* compare them */
}
```

Figure 1: An example of the two code fragments of WritableComparator.java. The one from the Stack Overflow post 22315734 (left) is outdated when compared to its latest version in the Hadoop code base (right). Its Apache v.2.0 license is also missing.

One participant in our survey expresses his/her concerns about this:

> "The real issue is less about the amount the code snippets on SO than it is about the staggeringly high number of software "professionals" that mindlessly use them without understanding what they're copying, and the only slightly less high number of would-be professionals that post snippets with built-in security issues. A related topic is beginners who post (at times dangerously) misleading tutorials online on topics they actually know very little about. Think PHP/MySQL tutorials written 10+ years after mysql_* functions were obsolete, or the recent regex tutorial that got posted the other day on HackerNew (https://news.ycombinator.com/item?id=14846506). They're also full of toxic code snippets."

Although this activity of online code cloning is well-known, there are only a few empirical studies on the topic [1], [3], [4], especially on finding the origins of the clones on Q&A websites. In this study, we tackle this challenge of establishing the existence of online code clones on Stack Overflow, investigate how they occur, and study the potential effects to software reusing them. Therefore, we mine Stack Overflow posts, detected online code clones, and analysed the clones to reveal "toxic code snippets."

Toxic code snippets mean code snippets that, after incorporating into software, degrade the software quality. Stack Overflow code snippets cloned from open source software or online sources can become toxic when they are (1) outdated, (2) violating their original software license, (3) exhibiting code smells, (4) containing faults, or (5) having security vulnerabilities. In this study, we focus on the first two forms of toxic code snippets, outdated code and license-violating code, as these two problems are still underexplored compared to code smells [75] and vulnerabilities [2], [20]. Moreover, Stack Overflow users also express their concerns about these two problems as shown in several discussion threads[2] on meta.stackexchange.com

about outdated answers and license of code on Stack Overflow. Outdated code snippets can be harmful since they are not up-to-date with their originals and may contain defects. Code snippets from open source projects usually fall under a specific software license, e.g., GNU General Public License (GPL). If they are cloned to Stack Overflow answers without the license, and then flow to other projects with conflicting licenses, legal issues may occur.

We would like to motivate the readers by giving two examples of toxic code snippets. The first example is an outdated and potentially license-violating online code clone in an answer to a Stack Overflow question regarding how to implement RawComparator in Hadoop[3]. Figure 1 shows—on the left—a code snippet embedded as a part of the accepted answer. The snippet shows how Hadoop implements the compare method in its WritableComparator class. The code snippet on the right shows another version of the same method, but at this time extracted from the latest version (as of October 3, 2017) of Hadoop. We can see that they both are highly similar except a line containing buffer.reset(null,0,0); which was added on November 21, 2014. The added line is intended for cleaning up the reference in the buffer variable and avoid excess heap usage (issue no. HADOOP-11323[4]). While this change has already been introduced into the compare method several years ago, the code example in Stack Overflow post is still unchanged. In addition, the original code snippet of WritableComparator class in Hadoop is distributed with Apache license version 2.0 while its cloned instance on Stack Overflow contains only the compare method and ignores its license statement on top of the file. There are two potential issues for this. First, the code snippet may appear to be under Stack Overflow's CC BY-SA 3.0 instead of its original Apache license. Second, if the code snippet is copied and incorporated into another software project with a conflicting license, a legal issue may arise.

The second motivating example of outdated online code clones with more disrupting changes than the first one can be found in an answer to a Stack Overflow question regarding how to format files sizes in a human-readable form. Figure 2 shows—on the left—a code snippet to perform the task from the StringUtils class in Hadoop.

---

2. Discussions about outdated answers and code license on Stack Overflow: meta.stackexchange.com/questions/131495, meta.stackexchange.com/questions/11705/, meta.stackexchange.com/questions/12527, meta.stackexchange.com/questions/25956, meta.stackoverflow.com/questions/321291.

3. http://stackoverflow.com/questions/22315734
4. https://issues.apache.org/jira/browse/HADOOP-11323

```java
/* Code in Stack Overflow post ID 801987 */
public static String humanReadableInt(long number) {
  long absNumber = Math.abs(number);
  double result = number;
  String suffix = "";
  if (absNumber < 1024) {
  } else if (absNumber < 1024 * 1024) {
    result = number / 1024.0;
    suffix = "k";
  } else if (absNumber < 1024 * 1024 * 1024) {
    result = number / (1024.0 * 1024);
    suffix = "m";
  } else {
    result = number / (1024.0 * 1024 * 1024);
    suffix = "g";
  }
  return oneDecimal.format(result) + suffix;
}
```

```java
/* StringUtils.java (2013-02-05) */
public static String humanReadableInt(long number) {
  return TraditionalBinaryPrefix.long2String(number,"",1);
}
```

Figure 2: An example of the two code fragments of `StringUtils.java`. The one from the Stack Overflow post 801987 (left) is outdated when compared to its latest version in the Hadoop code base (right). The toxic code snippet is outdated code and has race conditions.

The code snippet on the right shows another version of the same method, but at this time extracted from the latest version of Hadoop. We can see that they are entirely different. The `humanReadableInt` method is rewritten on February 5, 2013 to solve an issue of a race condition (issue no. HADOOP-9252[5]).

The two toxic code snippets in our examples have been posted on March 11, 2014 and April 9, 2009 respectively. They have already been viewed 259 and 2,886 times[6] at the time of writing this paper (October 3, 2017). Our calculation finds that there will be a new viewer of the first toxic snippet approximately every 5 days compared to almost every day for the second one. Considering the popularity of Stack Overflow, which has more than 50 million developers visiting each month[7], one toxic code snippet on Stack Overflow can spread and grow to hundred or thousand copies within only a year or two.

While research has mostly focused on reusing code snippets *from* Stack Overflow (e.g., [3], [33], [81]), fewer studies have been conducted on finding the origins of code examples copied *to* Stack Overflow and the awareness of Stack Overflow developers in doing so. Finding the origins of code examples reveals the problem of toxic code snippets caused by outdated code and software licensing violations. It is equally important to studying the effects of reusing Stack Overflow code snippets because it gives insights into the root cause of the problem and lays a foundation to an automatic technique to detect and report toxic code snippets on Stack Overflow to developers in the future.

This paper makes the following primary contributions:

1) **Awareness of Stack Overflow answerers and visitors to toxic code snippets:** We performed an online survey and collected answers from 201 highly-ranked Stack Overflow users and 87 Stack Overflow visitors. We found that the answerers cloned code snippets from open source projects to Stack Overflow answers. While Stack Overflow answerers are aware of their outdated code snippets, 19% of the participants rarely or never fix the code. 99% of the answerers never include a software license in their snippets and 69% never check for licensing conflicts. On the other hand, 66% of the Stack Overflow visitors experienced problems from reusing Stack Overflow code snippets, including outdated code. They are generally not aware of the CC BY-SA 3.0 license, and more than half of them never check for license compatibility when reusing Stack Overflow code snippets.

2) **A manual study of online code clones:** To empirically confirm the findings from the surveys, we used two clone detection tools to discover 2,289 similar code snippet pairs between 72,365 Java code snippets obtained from Stack Overflow's accepted answers and 111 Java open source projects from the curated Qualitas corpus [73]. We manually classified all of them.

3) **An investigation of toxic code snippets on Stack Overflow:** Our study shows that from the 2,289 online clones, at least 328 have been copied from open source projects or external online sources to Stack Overflow, potentially violating software licenses. For 153 of them, we found evidence that they have been copied from a specific open source project. 100 of them were found to be outdated, of which 10 were buggy code.

4) **An online code clone oracle:** The 2,289 manually investigated and validated online clone pairs are available for download[8] and can be used as a clone oracle.

## 2 EMPIRICAL STUDY

We performed an empirical study of online code clones between Stack Overflow and 111 Java open source projects

---

5. https://issues.apache.org/jira/browse/HADOOP-9252

6. The number of views is for the whole Stack Overflow post but we use it as a proxy of the number of views the accepted answer receives because the question and the answer of the two motivation examples have a short gap of posting time (within the same day and four days after).

7. Data as of 21 August 2017 from: https://stackoverflow.com

8. https://ucl-crest.github.io/cloverflow-web

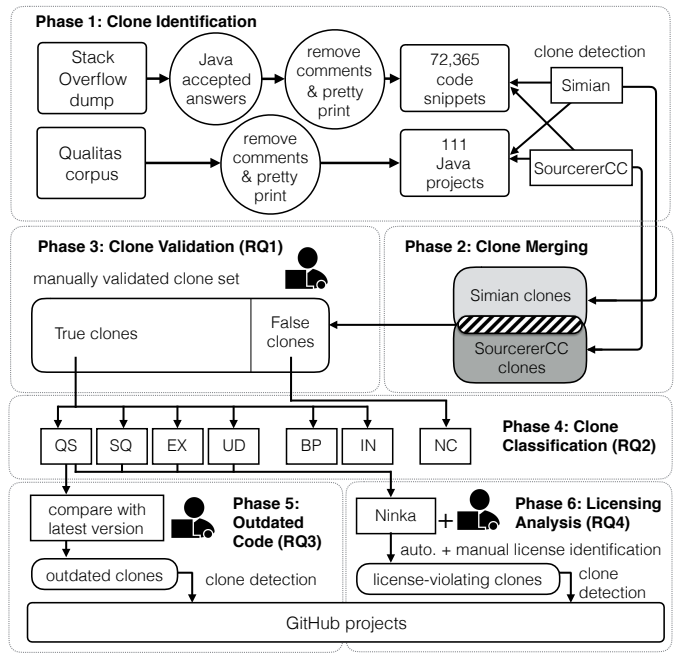

Figure 3: The Experimental framework

to answer the following research questions:

- **RQ1 (Stack Overflow answerers' and visitors' awareness to toxic code snippets):**
  *1) How often are Stack Overflow answerers aware of the outdated code and licensing conflicts when they answer a question on Stack Overflow?*
  *2) How often do Stack Overflow visitors experience the outdated code and licensing conflicts when they reuse code in an answer from Stack Overflow?*
  We surveyed 201 high-reputation Stack Overflow answerers and 87 Stack Overflow visitors to study their awareness of the two issues.
- **RQ2 (Online code clones):** *To what extent is source code cloned between Stack Overflow and open source projects?* We quantitatively measured the number of online code clones between Stack Overflow and open source projects to understand the scale of the problem.
- **RQ3 (Patterns of online code clones):** *How do online code clones occur?* We categorised online clones into seven categories allowing insights into how online code clones are created.
- **RQ4 (Outdated online code clones):** *Are online code clones up-to-date compared to their counterparts in the original projects?* We were interested in the outdated Stack Overflow code examples since users are potentially reusing them.
- **RQ5 (Software licensing violation):** *How often do license conflicts occur between Stack Overflow clones and their originals?* We investigated whether the reuse of online code clones can cause software developers to violate licenses.

To answer these five research questions, we perform two surveys and an empirical study to understand the developers' awareness of toxic code snippets on Stack Overflow and to empirically study the online code clones between Stack Overflow and open source projects, and their toxicity.

## 2.1 Stack Overflow Developers' Survey

We support our motivation of toxic code snippets on Stack Overflow and answer RQ1 by asking Stack Overflow users to take an online survey. The survey was used for assessing awareness of the developers on the two issues of outdated code and license-violating code snippets. We designed the survey using Google Forms by following the guidelines by Pfleeger and Kitchenham [35], [50]. The survey was completely anonymous, and the participants could decide to leave at any time. We created two versions of the survey: **the answerer survey** and **the visitor survey**. The answerer survey targeted the developers who were experienced Stack Overflow users and were highly active in answering questions. The visitor survey targeted the developers who searched for solutions and reused code from Stack Overflow answers.

**The answerer survey:** the survey contained 11 questions. There were 7 Likert's scale questions, 3 yes/no questions, and one open-ended question for additional comments. The first two questions were mandatory while the other 9 questions were shown to the participants based on their previous answers. The full survey can be found in our research note [56]. We selected the participants for the answerer survey based on their Stack Overflow reputation. On Stack Overflow, a user's reputation reflects how much the community trusts them. A user earns reputations when he or she receives upvotes for good questions and useful answers. Accepted answers receive more reputation score than questions and regular answers[9]. Thus, Stack Overflow reputation is an indicator of user's skills and their involvement in asking and answering questions on the site. In this study, we call Stack Overflow users who have a high reputation "Stack Overflow answerers." The participants were invited to take the survey via email addresses available on their public Stack Overflow and GitHub profiles. We selected the answerers based on the all-time reputation ranking[10]. The invited participants had a reputation from 963,731 (the highest) to 6,999, and we sent out 607 emails (excluding undelivered ones, e.g., due to illegal email addresses). The survey was open for participation for two months, from 25 July to 25 September 2017, before we collected and analysed the responses.

**The visitors' survey** The survey consists of 16 questions: 9 Likert's scale questions, 3 yes/no questions, 2 multiple-choice questions, and 2 open-ended questions. The first four questions are mandatory while the other 12 questions will be shown to the participants based on their previous answers. The survey collects information about the participant's software development experience, the importance of Stack Overflow, reasons for reusing Stack Overflow snippets, problems from Stack Overflow snippets, licensing of code on Stack Overflow, and additional feedback. The full survey can be found in our research note [56]. We adopted non-probability convenient sampling to invite participants for this survey. Participation in the survey requires experience of visiting Stack Overflow for solving programming tasks at least once. The participants were invited to take the survey via five channels: social media post (Facebook), blognone.com (a popular technology news and media community in Thailand), the University of Molise in Italy where the third author works, the `comp.lang.java.programmer` group, and the Software Engineering Facebook page. The survey was open for participation for 2 months from 25 July 2017 to 25 September 2017.

---

9. Stack Overflow Reputation: https://stackoverflow.com/help/whats-reputation
10. Stack Overflow Users (data as of 25 July 2017): https://stackoverflow.com/users?tab=Reputation&filter=all

### 2.2 Empirical Study of Online Code Clones

We support the motivation and confirm the findings in the surveys by performing an empirical study of online code clones between Stack Overflow answers and 111 Java open source projects. We designed the study in 6 phases as depicted in Figure 3 where we build different data sets to answer RQ2 to RQ5.

#### 2.2.1 Phase 1: Clone Identification

We rely on two source code data sets in this study: Java code snippets in answers on Stack Overflow and open source projects from the Qualitas corpus [73], as detailed next.

**Stack Overflow:** We extracted Java code snippets from a snapshot of a Stack Overflow dump[11] in January 2016. The data dump is in XML, and it contains information about posts (questions and answers). We were interested in code snippets embedded in posts which were located between `<code>...</code>` tags. A Stack Overflow thread contains a question and several answers. An answer can also be marked as an **accepted answer** by the questioner if the solution fixes his/her problem. We collected Java code snippets using two criteria. First, we only focused on code snippets in accepted answers. We chose the snippets in accepted answers because they actually solved the problems in the questions. Moreover, they are usually displayed just below the questions which makes them more likely to be reused than other answers. Second, we were only interested in code snippets of at least ten lines. Although the minimum clone size of six lines is usual in clone detection [8], [37], [79], we empirically found that snippets of six lines contain a large number of boiler-plate code of getters/setters, `equal` or `hashCode` methods, which are not interesting for the study. Each snippet was extracted from the dump and saved to a file. Moreover, we filtered out irrelevant code snippets that were part of the accepted answers but were not written in Java by using regular expressions and manual checking. Finally, we obtained 72,365 Java code snippets containing 1,840,581 lines[12] of Java source code. The median size of the snippets is 17 lines.

**Open source systems:** We selected the established **Qualitas** corpus [73]. It is a curated Java corpus that has been used in several software engineering studies [7], [48], [72], [76]. The projects in the corpus represent various domains of software systems ranging from programming languages to visualisation. We selected the release 20130901r of the

---

11. https://archive.org/details/stackexchange
12. Measured by cloc: https://github.com/AlDanial/cloc

Table 1: Stack Overflow and Qualitas datasets

| Data set | No. of files | SLOC |
| --- | --- | --- |
| Stack Overflow | 72,365 | 1,840,581 |
| Qualitas | 166,709 | 19,614,083 |

Qualitas corpus containing 112 Java open source projects. This release contains projects with releases no later than 1st September 2013. We intentionally chose an old corpus from 2013 since we are interested in online code clones in the direction from open source projects to Stack Overflow. The 20130901r snapshot provides Java code that is more than 2 years older than the Stack Overflow snapshot, which is sufficiently long for a number of code snippets to be copied onto Stack Overflow and also to observe if clones become outdated. Out of 112 Qualitas projects, there is one project, jre, that does not contain Java source code due to its licensing limitation [73] and is removed from the study. This resulted in 111 projects analysed in the study, for a total of 166,709 Java files containing 19,614,083 lines of code (see Table 1). The median project size is 60,667 lines of code.

**Clone Detection Tools:** We use clone detection to discover online code clones. There are a number of restrictions in terms of choosing the clone detection tools for this study. The main restriction is due to the nature of code snippets posted on Stack Overflow, as most of them are incomplete Java classes or methods. Hence, a detector must be flexible enough to process code snippets that are not compilable or not complete blocks. Moreover, since the amount of code that has to be processed is in a scale of millions line of code (as shown in Table 1), a clone detector must be scalable enough to report clones in a reasonable amount of time. We have tried 7 state-of-the-art clone detectors including Simian [66], SourcererCC [64], NiCad [13], [62], CCFinder [29], iClones [25], DECKARD [28], and PMD-CPD [51] against the Stack Overflow and Qualitas datasets. NiCad failed to parse 44,960 Stack Overflow snippets while PMD CPD failed to complete the execution due to lexical errors. iClones could complete its execution but skipped 367 snippets due to malformed blocks in Stack Overflow data sets. CCFinder reported 8 errors while processing the two data sets. Although Simian, SourcererCC, and DECKARD could successfully report clones, we decided to choose only Simian and SourcererCC due to their fast detection speed. Moreover, Simian and SourcererCC complement each other as SourcererCC's clone fragments are always confined to method boundaries while Simian's fragments are not.

**Simian** is a text-based clone detector that locates clones at line-level granularity and has been used extensively in several clone studies [11], [39], [40], [41], [42], [43], [45], [57], [58], [59], [79]. Furthermore, it offers normalisation of variable names and literals (strings and numbers) which enables Simian to detect literal clones (type-1) and parameterised clones (type-2) [8]. **SourcererCC** is a token-based clone detector which detects clones at either function- or block-level granularity. It can detect clones of type-1, -2 up to type-3 (clones with added and removed statements) and offer scalability against large code corpus [63], [64], [82].

We prepared the Java code in both datasets by removing comments and pretty-printing to increase the clone detection accuracy. Then, we deployed the two detectors to locate clones between the two datasets. For each Qualitas project, we ran the tools on the project's code and the entire Stack Overflow data. Due to incomplete code blocks and functions typically found in Stack Overflow snippets, the built-in SourcererCC's Java tokeniser could not parse 45,903 snippets, more than half of them. Nevertheless, the tool provides an option to plug in a customised tokeniser, so we developed a special Java tokeniser with assistance from the tool's creators. The customised tokeniser successfully processed all Stack Overflow snippets.

Table 2: Configurations of Simian and SourcererCC

| Tool | Configurations |
| --- | --- |
| Simian (*S*) | Threshold=10, ignoreStringCase, ignoreCharacterCase, ignoreModifiers |
| SourcererCC (*SCC*) | Functions, Minimum clone size=10, Similarity=80% |

Table 3: Number of online clones reported by Simian and SourcererCC

| Tool | Total clone pairs | Average clone size |
| --- | --- | --- |
| Simian | 721 | 16.61 |
| SourcererCC | 1,678 | 17.86 |

Simian did not provide an option to detect cross-project clones. Hence the Simian clone report was filtered to contain only clone pairs between Stack Overflow and Qualitas projects, removing all clone pairs within either Stack Overflow or Qualitas. SourcererCC can detect cross-project clones between two systems, so we did not filter the clones.

**Clone Detection Configuration:** We are aware of effects of configurations to clone detection results and the importance of searching for optimised configurations in empirical clone studies [57], [58], [60], [71], [78]. However, considering the massive size of the two datasets and the search space of at least 15 Simian and 3 SourcererCC parameters, we are unable to search for the best configurations of the tools. Thus, we decided to configure Simian and SourcererCC based on their established default configurations chosen by the tools' creators as depicted in Table 2. The two clone detectors complemented each other by having Simian detecting literal copies of code snippets (type-1) and SourcererCC detecting clones with renaming and added/deleted statements (type-2, type-3).

Nevertheless, we investigated a crucial parameter setting for clone detection: the minimum clone size threshold. Choosing a large threshold value can reduce the number of trivial clones (e.g., `equals`, `hashCode`, or getter and setter methods) and false clones in the analysis or the manual investigation phase [64], i.e., increasing precision. Nonetheless, it may create some false negatives. On the other hand, setting a low threshold results in a larger number of clone candidate pairs to look at, i.e., increasing recall, and a higher chance of getting false positives. Moreover, the large number of clone pairs hinders a full manual validation of the clones. Three threshold values, six, ten, and fifteen lines, were chosen for our investigation. We started our investigation by using a threshold value of six lines, a well-accepted minimum clone size in clone benchmark [8]. Simian reported 67,172 clone candidate pairs and SourcererCC reported 7,752 clone candidate pairs. We randomly sampled 382 pairs from the two sets for a manual check. This sample number was a statistically significant sample with a 95% confidence level and ±5% confidence interval. The first author investigated the sampled clone pairs and classified them into three groups: not clones, trivial clones (`equals`, `hashCode`, or getter and setter methods), and non-trivial clones. The manual check found 26 non-clone pairs, 322 trivial clone pairs, and 34 non-trivial clone pairs. Next, we increased the threshold to ten lines, another well-established minimum clone size for large-scale data sets [64], and retrieved 721 clone pairs from Simian and 1,678 clone pairs from SourcererCC. We randomly sampled and manually checked the same amount of 382 pairs and found 27 non-clone pairs, 253 trivial clone pairs, and 102 non-trivial clone pairs. Then, we increased the threshold further to fifteen lines and retrieved 196 clone pairs from Simian and 1,230 clone pairs from SourcererCC. The manual check of the 382 randomly sampled pairs revealed zero non-clone pairs, 298 trivial clone pairs, and 83 non-trivial clone pairs.

The findings from the three threshold values show that selecting the minimum clone size of ten lines was preferred over six and fifteen lines. First, it generated a fewer number of clone pairs than using six lines, which made the manual clone investigation feasible. Second, it preserved the highest number of non-trivial clone pairs. The number of online clone pairs reported using the minimum clone size of 10 lines are presented in Table 3. Simian reports 721 clone pairs while SourcererCC reports 1,678 clone pairs. The average clone size reported by Simian is 16.61 lines which is slightly smaller than SourcererCC (17.86 lines).

#### 2.2.2 Phase 2: Clone Merging

Clones from the two detectors can be duplicated. To avoid double-counting of the same clone pair, we adopted the idea of **clone agreement** which has been used in clone research studies [21], [60], [79] to merge clones from two data sets. Clone pairs agreed by both clone detection tools have a high likelihood to be duplicate and must be merged. To find agreement between two clone pairs reported by two different tools, we used the clone pair matching metric proposed by Bellon et al. [8]. Two clone pairs that have a large enough number of overlapping lines can be categorised as either a good-match or an ok-match pair with a confidence value between 0 and 1. Although good-match has a stronger agreement than ok-match, we choose the ok-match criterion as our clone merging method because it depends on clone containment and does not take clone size into account. Clone containment suits our online code clones from two tools, Simian (line-level) and SourcererCC (method-level), better because Simian's clone fragments can be smaller or bigger than a method while SourcererCC's clone fragments are always confined to a method boundary.

We follow Bellon's original definitions of ok-match [8], which are based on how much two clone fragments *CF* are

contained in each other:

$$contained(CF_1, CF_2) = \frac{|lines(CF_1) \cap lines(CF_2)|}{|lines(CF_1)|}$$

A clone pair $CP$ is formed by two clone fragments $CF_1$ and $CF_2$, i.e., $CP = (CF_1, CF_2)$ and the *ok-value* of two clone pairs is defined as

$$\begin{aligned}ok(CP_1, CP_2) = min(&max(contained(CP_1.CF_1, CP_2.CF_1),\\ &\quad contained(CP_2.CF_1, CP_1.CF_1)),\\ &max(contained(CP_1.CF_2, CP_2.CF_2),\\ &\quad contained(CP_2.CF_2, CP_1.CF_2)))\end{aligned}$$

Two clone pairs $CP_1$ and $CP_2$ are called an *ok-match(t)* iff, for threshold $t \in [0, 1]$ holds

$$ok(CP_1, CP_2) \geq t$$

The threshold $t$ is crucial for the ok-match because it affects the number of merged clone pairs. Setting a high $t$ value will result in a few ok-match clone pairs and duplicates of the same clone pairs (which are supposed to be merged) may appear in the merged clone set. On the other hand, setting a low $t$ value will result in many ok-match clone pairs, and some non-duplicate clone pairs may be accidentally merged by only a few matching lines. In order to get an optimal $t$ value, we did an analysis by choosing five $t$ values of 0.1, 0.3, 0.5, 0.7, 0.9 and studied the merged clone candidates. By setting $t = 0.7$ according to Bellon's study, we found 97 ok-match pairs reported. On the other hand, setting $t$ to 0.1, 0.3, 0.5, and 0.9 resulted in 111, 110, 110, and 94 ok-matched pairs respectively. Since the clone pairs of $t = 0.1$ were the superset of other sets, we manually checked all the 111 reported pairs. We found one false positive pair and 110 true positive pairs. By raising the $t$ to 0.3 and 0.5, we got rid of the false positive pair and still retained all the 110 true positive pairs. All the clone pairs of $t = 0.7$ (97) and $t = 0.9$ (94) were also true positives due to being a subset of $t = 0.5$. However, since there were fewer merged clone pairs, we ended up leaving some duplicates of the same clones in the final merged clone set. With this analysis, we can see that setting the threshold $t$ to 0.1 is too relaxed and results in having false positive ok-match pairs, while setting the $t$ to 0.7 or 0.9 is too strict. Thus, we decided to select the $t$ value at 0.5.

Using the ok-match criterion with the threshold $t$ of 0.5 similar to Bellon's study [8], we merge 721 clone pairs from Simian and 1,678 clone pairs from SourcererCC into a single set of 2,289 online clone pairs. There are 110 common clone pairs between the two clone sets as depicted in Figure 4. The low number of common clone pairs is due to SourcererCC reporting clones with method boundaries while Simian is purely line-based.

### 2.2.3 Phase 3-4: Validation and Classification

We used the 2,289 merged clone pairs for manual validation and online clone classification. The validation and classification of the pairs were done at the same time. The clone validation process (phase 3 in Figure 3) involves checking if a clone pair is a true positive or a false positive. Moreover, we are also interested in the patterns of code cloning so we can gain more insights into how these clones are created (phase 4 in Figure 3).

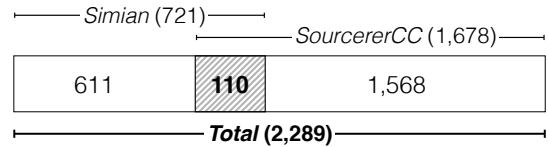

Figure 4: The result from clone merging using Bellon's ok-match criterion

Table 4: The seven patterns of online code cloning

| Patt. | Description |
|---|---|
| QS | Cloned from Qualitas project to Stack Overflow (Q → S) |
| SQ | Cloned from Stack Overflow to Qualitas project (S → Q) |
| EX | Cloned from an external source to Stack Overflow (X → S) |
| UD | Cloned from each other or from an external source outside the project (unknown) |
| BP | Boiler-plate or IDE auto-generated |
| IN | Inheritance, interface implementation |
| NC | Not clones |

**Manual investigation:** To mitigate the human error, we deployed two people in the manual clone investigation process. The first author, who is a research student working on clone detection research for three years, and the third author, who is a software engineering research student and familiar with code clones, took the role of the investigators performing a manual validation and classification of the merged clone pairs. The two investigators separately went through each clone pair candidate, looked at the clones, and decided if they are a true positive or a false positive and classified them into an appropriate pattern. After the validation, the results from the two investigators were compared. There were 338 (15%) conflicts between true and false clones (QS, SQ, EX, UD, BP, IN vs. NC). The investigators looked at each conflicting pair together and discussed until a consensus was made. Another 270 pairs (12%) were conflicts in the classification of the true clone pairs. Among these pairs, 145 conflicts were caused by one investigator being more careful than the other and being able to find evidence of copying while the other could not. Thus, resolving the conflicts lead to a better classification, i.e., from UD to QS or EX.

**The online cloning classification patterns:** We studied the eight patterns of cloning from Kapser et al. [30], [32] and performed a preliminary study to evaluate its applicability to our study. We tried to classify 697 online clone pairs from the reported clones in phase 1 using Kapser's cloning patterns. We found that Kapser's patterns are too broad for our study and a more suitable and fine-grained classification scheme is needed. After a preliminary study, we adopted one of Kapser's cloning patterns, *boiler-plate code*, and defined six new cloning patterns. The seven patterns include QS, SQ, EX, UD, BP, IN, and NC as presented in Table 4. Pattern QS (**Q**ualitas to **S**tack Overflow) represents clones that have evidence of being copied from a Qualitas project to Stack Overflow. The evidence of copying can be found in comments in the Qualitas source code or in the Stack Overflow post's contents. Pattern SQ (**S**tack Overflow to **Q**ualitas) is cloning, with evidence, in the opposite direction from Stack Overflow to a Qualitas project. Pattern EX (**Ex**ternal Sources) is cloning that has evidence of copying from a single or multiple external sources to Stack Overflow, and possibly also to a Qualitas project. Pattern UD (**U**nknown **D**irection) is cloning that creates identical or highly similar clones between Qualitas and Stack Overflow but where we could not find any attribution of copying. Pattern BP (**B**oiler-**P**late) represents clones containing boiler-plate. We define three cases of boiler-plate code and use in our classification as shown in Table 5. Our definition is specific to Java and more suitable to our study than the general definition in Kapser's [32]. Pattern IN (**In**heritance/Interface) is cloning by inheritance of the same super class or implementation of the same interface. These two activities usually result in similar overriding methods. The last pattern, NC (**N**ot **C**lones), represents false clone pairs. These are mainly false positive clones from the clone detectors such as similar `try-catch` statements.

Table 5: The definition of boiler-plate code

| Type | Description |
| --- | --- |
| API constraints | Similar code fragments are created because of a constraint by an API. For example, reading and writing to database using JDBC, reading and writing a file in Java. |
| Templating | An optimised or stable code fragment is reused multiple times. This also includes auto-generated code by IDE. |
| Design patterns | Java design patterns suggest a way of implementing similar pieces of code. For example, getters, setters, `equals`, `hashCode`, and `toString` method. |

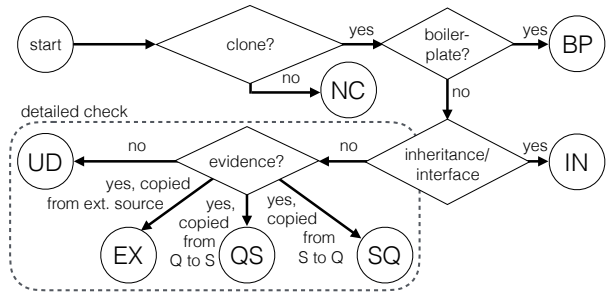

Figure 5: Online code clone classification process

The classification of the filtered online clone pairs followed the steps depicted in Figure 5. First, we look at a pair of clone fragments to see their similarity. If they are accidentally similar code fragments after code normalisation or false positive clones from the clone detection tools, we classify the pair into **NC**. If the two fragments are boiler-plate code, the pair is classified into **BP**. If they implement the same interface or inherited the same class and share similar overriding methods, the pair is classified into **IN**. If the pair is not **BP**, **IN**, or **NC**, we start a detailed investigation. We check the corresponding Stack Overflow post, read through it carefully and look for any evidence mentioning code copying. If evidence of copying has been found from a Qualitas project, the pair is classified in **QS**. In several occasions, we used extra information such as the questions' contents, the name of posters, and the tags to gain a better understanding. On the other hand, if the source code from the Qualitas project mentions copying from Stack Overflow, the pair is classified into **SQ**. If there is evidence of copying from an external source instead of a Qualitas project, the pair is classified as **EX**. Lastly, if there is no evidence of copying in any direction but the clone fragments are highly similar, we classify them into **UD**.

### 2.2.4 Phase 5: Outdated Clones

Outdated code occurs when a piece of code has been copied from its origin to another location, and later the original has been updated [80]. Usually, code clone detection is used to locate clone instances and update them to match with the originals [8]. However, online code clones are more difficult to detect than in regular software projects due to its large search space and a mix of natural and programming languages combined in the same post.

To search for outdated online code clones, we focused on the **QS** clone pairs that were cloned from Qualitas to Stack Overflow and compared them with their latest versions. We downloaded the latest version of the Qualitas projects from their repositories on October 3, 2017. For each **QS** online clone pair, we used the clone from Qualitas as a proxy. We searched for its latest version by the file name and located the cloned region in the file based on the method name or specific code statements. We then compared the Stack Overflow snippet to its latest version line-by-line to find if any change has been made to the source code. We also made sure that the changes did not come from the modifications made to the Stack Overflow snippets by the posters but from the updates in the projects themselves. When we found inconsistent lines between the two versions, we used `git blame` to see who modified those lines of code and the timestamps. We also read commit messages and investigated the issue tracking information if the code change is linked to an automatic issue tracking system, such as Jira or BugZilla to gain insights into the intent behind the change.

Lastly, we searched for the outdated code snippets in 130,719 GitHub projects to see how widespread is the outdated code in the wild. We mined GitHub based on the number of stars the projects received, which indicated their popularity. We relied on GitHub API to query the project metadata before cloning them. Since GitHub API returned only top 1,000 projects at a time for each query, we formulated the query to retrieve most starred projects based on their sizes. The project size range started from 1KB to 2MB with 1KB step size, and the last query is for all the remaining projects that were larger than 2MB. With this method, we retrieved the top 1,000 most starred projects for each project size. As a result, we cloned 130,719 GitHub projects ranging from 29,465 stars to 1 star. A clone detection was then performed between the outdated code snippets and the GitHub projects. We selected SourcererCC with the same settings (see Table 2) for this task because it could scale to a large-scale data set, while Simian could not. Finally, we analysed the clone reports and manually checked the clones.

### 2.2.5 Phase 6: Licensing Analysis

Software licensing plays an important role in software development. Violation of software licenses impacts software delivery and also leads to legal issues [68]. One can run into a licensing issue if one integrates third-party source code into their software without checking. A study by An et al. [3] reports 1,279 cases of potential license violations between 399 Android apps and Stack Overflow code snippets.

Table 6: Experience of Stack Overflow answerers

| Experience | Amount | Percent |
|---|---|---|
| Less than a year | 1 | 0.5% |
| 1 – 2 years | 1 | 0.5% |
| 3 – 5 years | 30 | 14.9% |
| 5 – 10 years | 58 | 28.9% |
| More than 10 years | 111 | 55.2% |

Table 7: Frequency of including code snippets in answers

| Include code snippets | Amount | Percent |
|---|---|---|
| Very Frequently (81–100% of the time) | 84 | 42% |
| Frequently (61–80% of the time) | 63 | 31% |
| Occasionally (41–60% of the time) | 40 | 20% |
| Rarely (21–40% of the time) | 11 | 6% |
| Very Rarely (1–20% of the time) | 2 | 1% |
| Never (0% of the time) | 1 | 1% |
| Total | 201 | 100% |

We analysed licensing conflicts of the online clones in the **QS**, **EX**, and **UD** set. The licenses were extracted by *Ninka*, an automatic license identification tool [23]. Since Ninka works at a file level, we report the findings based on Stack Overflow snippets and Qualitas source files instead of the clone pairs (duplicates were ignored). For the ones that could not be automatically identified by Ninka and have been reported as `SeeFile` or `Unknown`, we looked at them manually to see if any license can be found. For EX clone pairs that are cloned from external sources such as JDK or websites, we manually searched for the license of the original code. Lastly, we searched for occurrences of the license-conflicting online clones in GitHub projects using the same method as in the outdated clones.

## 3 RESULTS AND DISCUSSION

We use the two online surveys of Stack Overflow answerers and visitors to answer RQ1 and follow the 6 phases in the experimental framework (Figure 3) to answer the other four research questions. To answer RQ2, we rely on the number of manually validated true positive online clone pairs in phase 3. We use the results of the manual classification by the seven patterns of online code cloning to answer RQ3 (phase 4). For RQ4, we looked at the true positive clone pairs that are classified as clones from Qualitas to Stack Overflow and checked if they have been changed after cloning (phase 5). Similarly, for RQ5, we looked at the license of each clone in the pattern QS, EX, UD and checked for a possibility of license violation (phase 6).

### 3.1 RQ1: Stack Overflow Answerers' and Visitors' Awareness

#### 3.1.1 The Answerer Survey

We received 201 answers (33% response rate) from 607 emails we sent to the Stack Overflow answerers. The response rate was high considering other online surveys in software engineering [55]. We only present a summary of the survey answers in this paper, and the full analysis is available as a research note [56].

**General Information:** As shown in Table 6, the majority of the answerers are experienced developers with more than 10 years of experience (55.2%) or between 5 to 10 years (28.9%). They are active users and regularly answer questions. 49 participants (24%) answer questions on Stack Overflow every day.

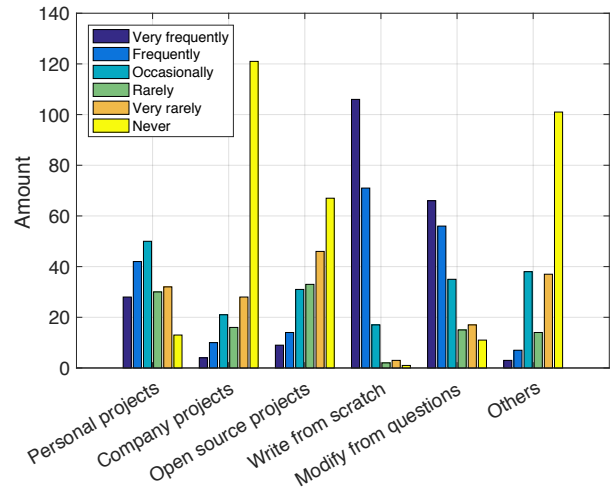

Figure 6: The sources of code snippets in Stack Overflow answers

**Code Snippets in Answers:** 84 and 63 answerers include code snippets in more than 80% and 60% of their answers respectively. Interestingly, there is one answerer who never include code snippet in his/her answers (Table 7).

We asked the remaining 200 participants for the origins of code snippets in their answers. We provided six locations including the answerer's personal projects, the answerer's company projects, open source projects, writing the code from scratch, copying and modifying the code from the question, and others (e.g., code that are copied from other questions or answers on Stack Overflow) and we asked them to rate how often they copied the code from these locations. The results are shown in Figure 6. Looking at the Very Frequently section, we can see that the answerers mainly write new code from scratch (106) or copy and modify the code snippets from the question for each answer (66), while fewer numbers are from other sources including their personal projects (28), their company projects (4), and open source projects (9). Although copying from open source projects is not the most popular choice, the answerers still rely on them sometimes. As shown in the figure, there are 14, 31, and 33 participants who frequently, occasionally, and rarely copied code snippets from open source projects.

*RQ 1.1 How often are Stack Overflow answerers aware of the outdated code and licensing conflicts when they answer a question on Stack Overflow?*

**Outdated Code Snippets:** We asked the answerers if they have ever been notified about outdated code in their answers. 111 participants selected *Yes* while the other 89 participants selected *No*. However, we found inconsistent results when we asked a follow-up question on the frequency of being notified. As displayed in Table 8, the number of participants who have *Never* been notified about outdated code snippets in their answers drops from 89 to 69.

Table 8: Notifications of outdated code snippets in answers

| Notified of outdated code | Amount | Percent |
| --- | --- | --- |
| Very frequently (81–100% of my answers) | 2 | 1% |
| Frequently (61–80% of my answers) | 1 | 0.5% |
| Occasionally (41–60% of my answers) | 9 | 4.5% |
| Rarely (21–40% of my answers) | 16 | 8% |
| Very rarely (1–20% of my answers) | 103 | 51.5% |
| Never (0% of my answers) | 69 | 34.5% |
| Total | 200 | 100% |

Table 9: Fixing of outdated code snippets in answers

| Fixing of outdated code | Amount | Percent |
| --- | --- | --- |
| Very frequently (81–100% of the cases) | 48 | 36.6% |
| Frequently (61–80% of the cases) | 27 | 20.6% |
| Occasionally (41–60% of the cases) | 30 | 22.9% |
| Rarely (21–40% of the cases) | 11 | 8.4% |
| Very rarely (1–20% of the cases) | 8 | 6.1% |
| Never (0% of the cases) | 7 | 5.3% |
| Total | 131 | 100.0% |

We found that although the answerers have been notified of outdated code in their answers, for 51.5% of them such notifications occur very rarely (only 1–20% of the answers). Only 3 participants reported that they were notified in more than 60% of their answers. This notification to the answerer can be done via several means, such as contacting the author directly or writing a comment saying that the answer is already outdated. The low percentage of outdated code notifications reflect the experience of high reputation answerers who accumulate the reputation for a long time. Due to the voting mechanism of Stack Overflow, high-reputation users usually provide high-quality answers to earn upvotes from other users. They are careful when posting code snippets in the answer to avoid problems and, vice versa, getting downvotes. It would be interesting to compare the findings to Stack Overflow answerers who are newer and have a lower reputation. However, we leave it to future work.

We then asked 131 participants who have been notified of their outdated code a follow-up question *"how frequently did you fix your outdated code on Stack Overflow?"*. The answers, depicted in Table 9, show that more than half of them (57.2%) very frequently or frequently fix the outdated code snippets. However, there are 19.8% of the answerers in both groups who rarely, very rarely, or never fix their outdated code.

Table 10: Inclusion of software license in answer

| Include license? | Amount |
| --- | --- |
| No. | 197 |
| Yes, in code comment | 1 |
| Yes, in text surrounding the code | 2 |
| Total | 200 |

Table 11: Checking for licensing conflicts with CC BY-SA 3.0

| Check license conflicts? | Amount | Percent |
| --- | --- | --- |
| Very Frequently (81–100% of the time) | 14 | 7% |
| Frequently (61–80% of the time) | 7 | 3.5% |
| Occasionally (41–60% of the time) | 10 | 5% |
| Rarely (21–40% of the time) | 16 | 8% |
| Very rarely (1–20% of the time) | 15 | 7.5% |
| Never (0% of the time) | 138 | 69% |
| Total | 200 | 100% |

**The License of Code Snippets:** Among the 200 answerers who include code snippets in their answers, 124 answerers are aware that Stack Overflow applies Creative Commons Attribution-ShareAlike 3.0 Unported (CC BY-SA 3.0) to content in the posts, including code snippets, while the rest (76) are not. Nevertheless, as shown in Table 10, almost all of them (197) reported that they did not include license statement in their code snippets due to several reasons. First, some answerers chose to post only their own code or code that was adapted from the question; hence they are automatically subjected to CC BY-SA 3.0. Second, they copied the code from company or open source projects that they knew were permitted to be publicly distributed. Third, some answerers believe that code snippets in their answers are too small to claim any intellectual property on them and fall under fair use [19].

While nobody explicitly includes a software license in their snippets, many users include a statement on their profile page that all their answers are under a specific license. For example, a Stack Overflow user includes the following text in his/her profile page.

> *All code posted by me on Stack Overflow should be considered public domain without copyright. For countries where public domain is not applicable, I hereby grant everyone the right to modify, use and redistribute any code posted by me on Stack Overflow for any purpose. It is provided "as-is" without warranty of any kind.*

Many users either declare their snippets to be public domain, or they grant additional licenses, e.g., Apache 2.0 or MIT/Expat.

We asked the answerers a follow-up question of how frequently they checked for a conflict between software license of the code snippets they copied to their answers and Stack Overflow's CC BY-SA 3.0. As shown in Table 11, approximately 69% of answerers did not perform the checking. Nonetheless, there are about 10.5% of the answerers who very frequently or frequently check for licensing conflicts when they copy code snippets to Stack Overflow.

> To answer RQ 1.1, although most of the Stack Overflow answerers are aware that their code can be outdated, 51.5% of the answerers were very rarely notified and 35.5% have never been notified of outdated code in the answers. After being notified, 19.8% of them rarely or never fix the outdated code. 124 answerers out of 200 (62%) are aware of Stack Overflow's CC BY-SA 3.0 license applied to code snippets in questions and answers. However, only 3 answerers explicitly include software license in their answers. Some answerers choose to include the license in their profile page instead. 69% of the answerers never check for licensing conflicts between their copied code snippets and Stack Overflow's CC BY-SA 3.0.

**Open Comments:** We also invited the participants to give a free-form comment regarding their concerns about answering Stack Overflow with code snippets. Besides the one we present earlier in the introduction, these are interesting comments we received.

1) *"When I copy code it's usually short enough to be considered "fair use" but I am not a lawyer or copyright expert so some guidance from Stack Overflow would be helpful. I'd also like the ability to flag/review questions that violate these guidelines."*
2) *"My only concern, albeit minor, is that I know people blindly copy my code without even understanding what the code does."*
3) *"The main problem for me/us is outdated code, esp. as old answers have high google rank so that is what people see first, then try and fail. Thats why we're moving more and more of those examples to knowledge base and docs and rather link to those."*
4) *"Lot of the answers are from hobbyist so the quality is poor. Usually they are hacks or workarounds (even MY best answer on Stack Overflow is a workaround)."*

The comments highlight that Stack Overflow users are unsure about the legal implications of copying code, that code is copied without understanding it, and that the quality of code on Stack Overflow is often low.

### 3.1.2 The Visitor Survey

We received answers from 89 participants. Two participants never copy code from Stack Overflow, so we analysed the answers of the remaining 87 participants. We only present a summary of the survey answers in this paper, and the full analysis is available as a research note [56].

**General Information:** Twenty-four (27%) and twenty-one (24%) participants have over 10 years and 5–10 years of experience respectively. There are 19 participants (21%) who have 3–5 years, 18 (20%) who have 1-2 years, and 7 (8%) participants who have less than a year of programming experience.

**The Choice for programming solutions:** Stack Overflow is ranked higher than official documentation, online repositories, and books as the resource to look for programming solutions. Developers rely on Stack Overflow answers because they are easy to search for on the web. Moreover, 64% of the participants reuse code snippets from Stack Overflow at least once a week. They copy code from Stack Overflow because they can be found easily from a search engine, solve similar problems to their problems, provide helpful context, and offer voting mechanism and accepted answers.

*RQ 1.2 How often do Stack Overflow visitors experience the outdated code and licensing conflicts when they reuse code in an answer from Stack Overflow?*

The survey results show that 57 out of 87 Stack Overflow visitors encountered a problem from reusing Stack Overflow code snippets. Ten participants experienced problems for more than 80% of the copied snippets, and sixteen participants faced problems for 40–60% of the reused code. As shown in Table 12, the problems ranked by frequency include mismatched solutions (40), outdated solutions (39), incorrect solutions (28), and buggy code (1). Sixty-three percent of the participants never report the problems back to Stack Overflow (Table 13). The ways of reporting the problems (22 answers) included down-voting the answer containing the problematic code snippet (8), writing a comment saying that the code has problems (10), contacting the answerers regarding the problems directly (2), and posting a better snippet as new answer on same topic (2).

In addition, 74 out of the 87 (85%) participants are not aware of Stack Overflow CC BY-SA 3.0 license, and 62% never give attributions to the Stack Overflow posts they copied the code snippets from. As shown in Table 14, we found that 66% of the visitors never check for software licensing conflicts between Stack Overflow code snippets and their projects. Interestingly, 9% of the participants encountered legal issues. Due to the anonymity of the survey, we could not investigate further regarding the legal issues that the participants faced from using Stack Overflow code snippets. To the best of our knowledge, a study of legal problems from reusing Stack Overflow code snippets has never been done before and will be our future work.

To answer RQ 1.2, Stack Overflow visitors experienced several issues from Stack Overflow answers including outdated code. 85% of them are not aware of CC BY-SA 3.0 license enforced by Stack Overflow and 66% never check for license conflicts when reusing code snippets.

Table 12: Problems from Stack Overflow code snippets

| Problem | Amount |
|---|---|
| Mismatched solutions | 40 |
| Outdated solutions | 39 |
| Incorrect solutions | 28 |
| Buggy code | 1 |

Table 13: Frequency of reporting the problems to Stack Overflow posts

| Report? | Amount | Percent |
|---|---|---|
| Very Frequently (81–100% of the time) | 1 | 1.8% |
| Frequently (61–80% of the time) | 1 | 1.8% |
| Occasionally (41–60% of problematic snippets) | 3 | 5.3% |
| Rarely (21–40% of problematic snippets) | 8 | 14.0% |
| Very rarely (1–20% of problematic snippets) | 8 | 14.0% |
| Never (0% of problematic snippets) | 36 | 63.2% |
| Total | 57 | 100% |

Table 14: Check for licensing conflicts before using Stack Overflow snippets

| License check? | Amount | Percent |
|---|---|---|
| Very frequently (81–100% of the time) | 0 | 0.0% |
| Frequently (61–80% of the time) | 7 | 8.1% |
| Occasionally (41–60% of the time) | 6 | 6.9% |
| Rarely (21–40% of the time) | 6 | 6.9% |
| Very rarely (1–20% of the time) | 11 | 12.6% |
| Never (0% of the time) | 57 | 65.5% |
| Total | 87 | 100% |

Table 15: Investigated online clone pairs and corresponding snippets and Qualitas projects

| Set | Pairs | Snippets | Projects | Cloned ratio |
|---|---|---|---|---|
| Reported clones | 2,289 | 460 | 59 | 53.28% |
| TP from manual validation | 2,063 | 443 | 59 | 54.09% |

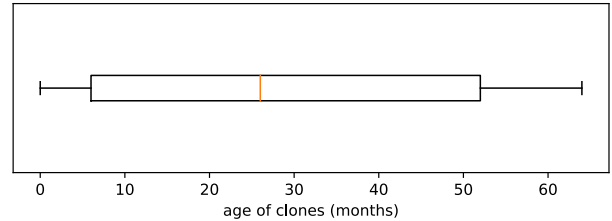

Figure 7: Age of QS online code clones.

## 3.2 RQ2: Online Code Clones

*To what extent is source code cloned between Stack Overflow and open source projects?*

The statistics on clones obtained from the merged clone data set are presented in Table 15. Simian and SourcererCC reported clones in 460 snippets, approximately 0.6% of the 72,365 Stack Overflow snippets, associated with 59 Qualitas projects. For the cloned Stack Overflow snippets, the average ratio of cloned code is 53.28% (i.e., the number of cloned lines of the cloned Stack Overflow snippet).

Lastly, during the manual investigation of 2,289 clone pairs, we identified 226 pairs as not clones (NC), i.e., false positives. After removing them, the set still contains 2,063 true positive clone pairs between 443 Stack Overflow snippets and 59 Qualitas projects. The average cloned ratio for the true positive clone pairs is 54.09%.

To answer RQ2, we found 2,063 manually confirmed clone pairs between 443 Stack Overflow code snippets and 59 Qualitas proejcts.

## 3.3 RQ3: Patterns of Online Code Cloning

*How do online code clones occur?*

We performed a manual classification of the 2,289 merged clone pairs by following the classification process in Figure 5. The classification results are shown in Table 16 and explained in the following.

**QS: Qualitas → Stack Overflow.** We found 247 online clone pairs with evidence of cloning from Qualitas projects to Stack Overflow. However, we observed that, in some cases, a cloned code snippet on Stack Overflow matched with more than one code snippet in Qualitas projects because of code cloning inside Qualitas projects themselves. To avoid double counting of such online clones, we consolidated multiple clone pairs having the same Stack Overflow snippet, starting line, and ending line into a single clone pair. We finally obtained 153 QS pairs (Table 16) having unique Stack Overflow code snippets and associated with 23 Qualitas projects listed in Table 17. The most cloned project is hibernate with 23 clone pairs, followed by eclipse (21 pairs), jung2 (19 pairs), spring (17 pairs), and jfreechart (13 pairs). The clones are used as examples and are very similar to their original Qualitas code with limited modifications. Most of them have a statement in the Stack Overflow post saying that the code is "copied," "borrowed" or "modified" from a specific file or class in a Qualitas project. For example, according to the motivating example in Figure 1, we found evidence in the Stack Overflow Post 22315734 saying that *"Actually, you can learn how to compare in Hadoop from WritableComparator. Here is an example that borrows some ideas from it."*

We analysed the time it took for the clone to appear from Qualitas projects to Stack Overflow. The clone ages were calculated by counting the number of months between the date of each Qualitas project and the date the answer was posted on Stack Overflow as shown in Figure 7. We found that, on average, it took the clones around 2 years since they appeared in Qualitas projects to appear on Stack Overflow answers. Some of the clones appeared on Stack Overflow almost at the same time as the original, while the oldest clones took around 5 years.

**SQ: Stack Overflow → Qualitas.** We found one pair with evidence of cloning from Stack Overflow post ID 698283 to POIUtils.java in jstock project. The user who asked the question on Stack Overflow is an author of jstock. The question is about determining the right method to call among 7 overloading methods of setCellValue during runtime. We could not find evidence of copying or attribution to Stack Overflow in jstock. However, considering that the 25 lines of code of findMethodToInvoke method depicted in Figure 8 in Stack Overflow is very similar to the code in jstock including comments, it is almost certain that the two code snippets form a clone pair. In addition, the Stack Overflow answer was posted on March 30, 2009, while the code in POIUtils class in jstock was committed to GitHub on the next day of March 31, 2009.

This very low number of SQ clone pair is very likely due to the age of the Qualitas corpus as another study [3] showed the presence of clones from Stack Overflow in newer open source data sets. This is expected and comes from our experimental design since we are more interested in cloning from Qualitas to Stack Overflow.

**EX: External Sources.** We found 197 clone pairs from external sources to Stack Overflow. After consolidating duplicated SO snippets due to multiple intra-clone instances in Qualitas, we obtained 109 EX pairs. We found evidence of copying from an external source to both Stack Overflow and Qualitas in 49 pairs. Each of the pairs contains statement(s) pointing to the original external location of the cloned code in Qualitas and Stack Overflow. Besides, we found evidence of copying from an external source to Stack Overflow, but not in Qualitas, in 60 pairs. Our analysis shows that the external sources fall into six groups as displayed in Figure 9. There are 63 EX online clone pairs copied

Table 16: Classifications of online clone pairs.

| Set | QS | SQ | EX | UD | BP | IN | NC | Total |
|---|---|---|---|---|---|---|---|---|
| Before consolidation | 247 | 1 | 197 | 107 | 1,495 | 16 | 226 | 2,289 |
| After consolidation | 153 | 1 | 109 | 65 | 216 | 9 | 53 | 606 |

Table 17: Qualitas projects associated with QS and UD online clone pairs

| QS | | UD | |
|---|---|---|---|
| Project | Pairs | Project | Pairs |
| hibernate | 23 | netbeans | 11 |
| eclipse | 21 | eclipse | 8 |
| jung2 | 19 | jstock | 5 |
| spring | 17 | compiere | 5 |
| jfreechart | 13 | ireport | 4 |
| hadoop | 10 | jmeter | 4 |
| tomcat | 8 | jung2 | 3 |
| log4j | 8 | jhotdraw | 3 |
| struts2 | 5 | c-jdbc | 3 |
| weka | 4 | log4j | 3 |
| lucene | 4 | wct | 2 |
| poi | 3 | hibernate | 2 |
| junit | 3 | tomcat | 2 |
| jstock | 2 | spring | 1 |
| jgraph | 2 | rssowl | 1 |
| jboss | 2 | mvnforum | 1 |
| jasperreports | 2 | jfreechart | 1 |
| compiere | 2 | jboss | 1 |
| jgrapht | 1 | hadoop | 1 |
| itext | 1 | geotools | 1 |
| c-jdbc | 1 | freemind | 1 |
| ant | 1 | findbugs | 1 |
| antlr4 | 1 | cayenne | 1 |

```
private Method findMethodToInvoke(Object test) {
  Method method = parameterTypeMap.get(test.getClass());
  if (method != null) {
    return method;
  }

  // Look for superclasses
  Class<?> x = test.getClass().getSuperclass();
  while (x != null && x != Object.class) {
    method = parameterTypeMap.get(x);
    if (method != null) {
      return method;
    }
    x = x.getSuperclass();
  }

  // Look for interfaces
  for (Class<?> i : test.getClass().getInterfaces()) {
    method = parameterTypeMap.get(i);
    if (method != null) {
      return method;
    }
  }
  return null;
}
```

Figure 8: The `findMethodToInvoke` that is found to be copied from Stack Overflow post 698283 to `POIUtils` class in jstock.

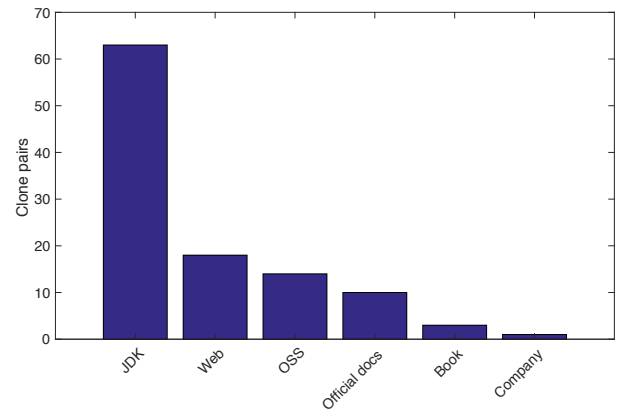

Figure 9: Original sources of EX clone pairs

from source code of Java SDK (e.g., java.util, javax.swing, javax.servlet), 18 pairs from websites, 14 pairs from open source systems not in Qualitas (e.g., Mozilla Rhino), 10 pairs from Java official documentation by Sun Microsystems or Oracle, 3 pairs from books, and 1 pair from a company project. For example, Stack Overflow Post 9549009 contains a code comment saying *"Copied shamelessly from org.bouncycastle.crypto.generators.PKCS5S2ParametersGenerator"* which is an open source project not included in the Qualitas corpus. Post 92962 includes a `VerticalLabelUI` class with a license statement showing that it is developed by a private company called Sapient. Post 12879764 has a text saying *"Code modified and cleaned from the original at Filthy Rich Clients."* which is a book for developing animated and graphical effects for desktop Java applications. Another example is a copy of the code from a website in post 15260207. The text surrounding source code reads *"I basically stole this from the web and modified it slightly... You can see the original post here (http://www.java2s.com/Code/Java/Swing-JFC/DragListDemo.htm)."*. Interestingly, the code is actually a copy from Sun Microsystems.

These findings complement a study of clones between software projects [71]. We found that cloning can also happen among different sources on the Internet just like software projects. There are 18 clone pairs that originated from programming websites including www.java2s.com and exampledepot.com. Moreover, there is one snippet which comes from a research website. We found that a snippet to generate graphical *Perlin noise* is copied from NYU Media Research Lab (http://mrl.nyu.edu/~perlin/noise/) website and is used in both Stack Overflow answer and the aoi project with attribution.

**UD: Unknown Direction.** We found 107 online clone pairs, reduced to 65 pairs after consolidating the clones, with no evidence of cloning between Qualitas and Stack Overflow but with a high code similarity that suggests cloning. The most cloned project is netbeans with 11 clone pairs. Most of the clones are a large chunk of code handling GUI components. Although these GUI clones might be auto-generated by IDEs, we did not find any evidence. The second most cloned project is eclipse (8 pairs), followed by jstock (5 pairs), a free stock market software, and compiere, a customer relationship management (CRM) system.

**BP: Boiler-Plate.** There were a large amount of boiler-plate clone pairs found in this study. We observed 1,495 such clone pairs and 216 after consolidation. The BP clone pairs account for 65% of all clone pairs we classified. The majority of them are `equals()` methods.

**IN: Inheritance/interface.** There were 16 clone pairs, 9 pairs after consolidation, found to be similar because they implement the same interface or inherit from the same class. An example is the two implementations of a custom data type which implements `UserType`. They share similar `@Override` methods of `deepCopy`, `isMutable`, `assemble`, `disassemble`, and `replace`.

**NC: Not Clones.** There were 226 non-clone pairs, 53 after consolidation. Mainly, they are false positive clones caused by code normalisation and false type-3 clones from SourcererCC. Examples of the NC clone instances include `finally` or `try-catch` clauses that were accidentally the same due to their tiny sizes, and similar `switch-case` statements.

> To answer RQ3, we found 153 pairs with strong evidences to be cloned from 23 Qualitas projects to Stack Overflow, 1 pair was cloned from Stack Overflow to Qualitas, and 109 pairs were found to be cloned to Stack Overflow from external sources. However, the largest amount of the clone pairs between Stack Overflow and Qualitas projects are boiler-plate code (216), followed by 65 clone pairs with no evidence that the code has actually been copied, and 9 pairs of clones due to implementing the same interface or inheriting the same class.

### 3.4 RQ4: Outdated Online Code Clones

*Are online code clones up-to-date compared to their counterparts in the original projects?*

We discovered 100 outdated online clone pairs out of 153 pairs. As shown in Figure 10, hibernate has the highest number of 19 outdated pairs, followed by 14 from spring, 13 from eclipse, and 9 from hadoop. Besides the two examples of outdated code in `WritableComparator` and `StringUtils` class from hadoop shown in Figure 1 and Figure 2, we also found a few outdated code elements which contained a large number of modifications. For example, the code snippet in Stack Overflow post 23520731 is a copy of `SchemaUpdate.java` in hibernate. The code has been heavily modified on February 5, 2016.

We analysed code modifications which made Stack Overflow code outdated by going through commits and git blame information. The six code modification types found in the 100 outdated online clone pairs are summarised in Table 18. They include statement addition, statement modification, statement removal, method rewriting, API change (changing in method signature), and file deletion. We occasionally found multiple code modifications applied to one clone pair at the same time but at a different location. The most often code change found is statement modification (50 occurrences), followed by statement addition (28 occurrences), statement removal (18 occurrences), change of method signature, i.e., API change (16 occurrences), and method rewriting (15 occurrences). Moreover, in the 100 outdated pairs, we found 14 "dead" snippets. These snippets cannot be located in the latest version of the projects. For example, the snippet in Stack Overflow post 3758110, a copy of `DefaultAnnotationHandlerMapping.java` in spring, was deleted in the commit `02a4473c62d8240837bec297f0a1f3cb67ef8a7b` on January 20, 2012, two years after it was posted.

Table 18: Six code modification types found when comparing the outdated clone pairs to their latest versions

| Modification | Occurrences |
| --- | --- |
| Statement modification | 50 |
| Statement addition | 28 |
| Statement removal | 18 |
| Method signature change | 16 |
| Method rewriting | 15 |
| File deletion | 14 |

Moreover, using the information in git commit messages, we can associate each change to its respective issues in an issue tracking system, such as Bugzilla or Jira. We found that in 58 cases, the cloned code snippets on Stack Overflow were changed because of a request in the issue tracking system. Since issue tracking systems are also used, besides bug reports, for feature request and feature enhancements, having an issue tracking ID can reflect that most of the changes are important and not only a superficial fix such as code formatting. The intent behind the changes are grouped into six categories as shown in Table 20. Enhancement is the majority intent accounting for 65 of the 100 outdated code (65%). Next is code deprecation (15%), which represents the code snippets that are outdated due to the use of deprecated functions or APIs. The code snippets that we analysed contained a few deprecated code statements while the newest version (based on the time we did the analysis) of the same code snippets no longer contain the deprecated part. There were 10 outdated code snippets (10%) caused by bug fixing. They contained buggy code statements, and were later fixed in the newest version. The rest of the changes are because of code refactoring (6%), changing coding style (3%), and the data format change (2%). Not all outdated code is toxic. However, the 10 buggy and outdated code snippets we found are toxic and are harmful to reuse.

Table 19 shows examples of the outdated online clones on Stack Overflow. The table displays information about the clones from both Stack Overflow and Qualitas side including the dates. We summarise the changes that make the clones outdated into three types, modified/added/deleted statements (*S*), file deletion (*D*), and method rewriting (*R*), along with the issue tracking number and the date of the change. The complete set of 100 outdated online clones can be found on the study website.

We performed a detailed investigation of the 100 outdated answers on Stack Overflow, on May 6, 2018, ap-

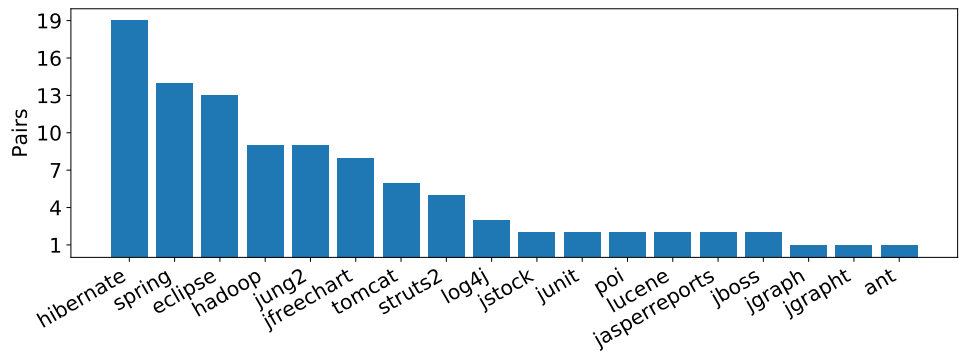

Figure 10: Outdated QS online clone pairs group by projects

Table 19: Examples of the outdated QS online clones

| Stack Overflow | | Qualitas | | | | | | Changes | | |
|---|---|---|---|---|---|---|---|---|---|---|
| Post | Date | Project | Ver. | File | Start | End | Date | Issue ID | Type* | Date |
| 2513183 | 25-Mar-10 | eclipse | 4.3 | GenerateToStringAction.java | 113 | 166 | 5-Jun-13 | Bug 439874 | S | 17-Mar-15 |
| 22315734 | 11-Mar-14 | hadoop | 1.0.0 | WritableComparator.java | 44 | 54 | 25-Aug-11 | HADOOP-11323 | S | 20-Nov-14 |
| 23520731 | 7-May-14 | hibernate | 4.2.2 | SchemaUpdate.java | 115 | 168 | 22-May-13 | HHH-10458 | S | 5-Feb-16 |
| 18232672 | 14-Aug-13 | log4j | 1.2.16 | SMTPappender.java | 207 | 228 | 31-Mar-10 | Bug 44644 | R | 18-Oct-08 |
| 17697173 | 17-Jul-13 | lucene | 4.3.0 | SlowSynonymFilterFactory.java | 38 | 52 | 6-Apr-13 | LUCENE-4095 | D | 31-May-12 |
| 21734562 | 12-Feb-14 | tomcat | 7.0.2 | FormAuthenticator.java | 51 | 61 | 4-Aug-10 | BZ 59823 | R | 4-Aug-16 |
| 12593810 | 26-Sep-12 | poi | 3.6 | WorkbookFactory.java | 49 | 60 | 7-Dec-09 | 57593 | R | 30-Apr-15 |
| 8037824 | 7-Nov-11 | jasperreports | 3.7.4 | JRVerifier.java | 1221 | 1240 | 31-May-10 | N/A | D | 20-May-11 |
| 3758110 | 21-Sep-10 | spring | 3.0.5 | DefaultAnnotationHandlerMapping.java | 78 | 92 | 20-Oct-10 | SPR-14129 | D | 20-Jan-12 |
| 14019840 | 24-Dec-12 | struts | 2.2.1 | DefaultActionMapper.java | 273 | 288 | 17-Jul-10 | WW-4225 | S | 18-Oct-13 |

* S: modified/added/deleted statements, D: file has been deleted, R: method has been rewritten completely

Table 20: Intents of code changes in the 100 outdated code snippets

| Intent | Detail | Amount |
|---|---|---|
| Enhancement | Add or update existing features | 64 |
| Deprecation | Delete dead/deprecated code | 15 |
| Bug | Fix bugs | 10 |
| Refactoring | Refactor code for better design | 6 |
| Coding style | Update to a new style guideline | 3 |
| Data change | Changes in the data format | 2 |

| Clones | Amount |
|---|---|
| *Found in Qualitas GitHub repos* | 13 |
| *Found in other project repos* | |
| Exact copy (outdated) | 47 |
| Non-exact copy | 32 |
| Total | 102 |

Table 21: Clones of the 100 Stack Overflow outdated code snippets in 131,703 GitHub projects

proximately two years after the snapshot we analysed was created to look for any changes, warnings, or mitigations made to the outdated code snippets. We investigated the answers on three aspects: newer answers, higher-voted answers, and comments on the outdated answers.[13] We found 34 posts which contained newer answers and 5 posts which contained answers with a higher number of votes than the outdated answers. However, 99 of the 100 outdated answers were still marked as accepted answers. For the comments, we check if there is any comment to mitigate or point out the toxicity of the outdated code snippets. We found that, out of 100 answers, 6 answers had a comment saying the code in the answer is outdated or containing issues, such as "spring 3.1 stuff", "...tried this but having connect exception –

---

13. We also found that the Stack Overflow question ID 22262310, which has an outdated answer from `WritableComparator.java` (Figure 1), has a negative score of -1. It is interesting to see if the negative score for the question impacts both the number of views and the trust that people place in the answer to a down-voted question. However, it is out of the scope of this study so we leave it as future work.

`javax.mail.MessagingException`: *Could not connect to SMTP host:* `smtp.gmail.com`, *port: 465"*, *"You should add a* `buffer.reset(null, 0, 0);` *at the end of the try block to avoid excess heap usage (issue no. HADOOP-11323)"* or *".. I do not have experience with new versions of hibernate for a long time. But previously without clean you could receive some unexpected results. So I suggest to try different approaches or even check latest documentation"*. The 6 outdated code snippets were still not fixed, but the comments themselves may help to warn some of the Stack Overflow users.

Then, we performed code clone detection between the 100 outdated code snippets and 130,719 GitHub projects. We found 102 cloned candidates, which were associated with 30 outdated code snippets, appearing in 68 GitHub projects and manually investigated all of them. Out of the 102 cloned snippets, 13 cloned snippets matched with themselves because some of the Qualitas projects also appear on GitHub. For other projects besides the Qualitas projects, 32 cloned snippets were not exactly the same (e.g., they contained additional code modifications made by the projects' developers, or they were copied from another source with a slightly different code). 47 cloned snippets were the same as the outdated code snippets, which of 12 were buggy. Two cloned snippets gave attributions to Stack Overflow. The attributions pointed to different posts than the ones we found but containing the same code in the answers[14]. 32 cloned snippets were very likely to be a file-level clone from its respective original project (e.g., JFreeChart, JUnit, Log4J, Hadoop) based on their license header and the Javadoc comments. 13 cloned snippets did not have any hints or evidence of copying.

Interestingly, we discovered that the buggy version of the `humanReadableInt()` method from Hadoop appeared in two popular Java projects: deeplearning4j (8,830 stars and 4,223 forks) and Apache Hive (1,854 stars and 1,946 forks). Due to the lack of evidence, we could not conclude how this method, which is the same as the toxic code snippet we found on Stack Overflow, appears in the two projects. It is possible that the developers retrieved them from Stack Overflow, other websites, or from Hadoop code base directly. Nevertheless, we reported them to the developers of the two projects regarding the issue. We created a bug report for each project (deeplearning4j #4694[15] and HIVE-18929[16]) and communicated with the developers of the projects by describing the problem of race condition in the outdated version of the `humanReadableInt()` method and proposed a fix by using the newest version of the method in Hadoop. The issue has been fixed in both projects. The developers of deeplearning4j agreed that the method was problematic and decided to implement their own fix due to a concern of a potential software licensing conflict caused by copying the fix directly from the Hadoop code base. The developers of Apache Hive found that the method was not used anywhere in the project, so they decided to delete it.

Although we did not find strong evidence of the outdated code snippets in GitHub projects, it would still be useful if Stack Overflow implements a flagging of outdated answers. The outdated online code clones cause problems ranging from uncompilable code (due to modifications and different API usage in the outdated code) to the introduction of vulnerabilities to the software [80]. An outdated code with a subtle change (e.g., Figure 1) may be copied and reused without awareness from developers. Moreover, an outdated code with a defect (e.g., a race condition problem in Figure 2) is clearly harmful to be reused. Although Stack Overflow has a voting mechanism that may mitigate this issue, the accepted answer may still be used by naive developers who copy and reuse the outdated code.

> For RQ4, our results show that 66% (101) of QS clone pairs on Stack Overflow are outdated. 86 pairs differ from their newest versions by modifications applied to variable names or method names, added or deleted statements, to a fully rewritten code with new method signatures. 14 pairs are dead snippets. 47 outdated code snippets

Table 22: License mapping of online clones (file-level)

| Type | Qualitas | Stack Overflow (CC BY-NC-SA) | QS | EX | UD |
|---|---|---|---|---|---|
| Compatible | Apache-2 | Apache-2 | 1 | | |
| | EPLv1 | EPLv1 | 2 | | 1 |
| | Proprietary | Proprietary | | 2 | |
| | Sun Microsystems | Sun Microsystems | | 3 | |
| | No license | No license | 20 | 9 | 2 |
| | No license | CC BY-SA 3.0 | | 1 | |
| Total | | | 23 | 15 | 3 |
| Incompat. | AGPLv3/3+ | No license | 1 | | 4 |
| | Apache-2 | No license | 46 | 14 | 12 |
| | BSD/BSD3 | No license | 4 | | 1 |
| | CDDL or GPLv2 | No license | | | 6 |
| | EPLv1 | No license | 10 | | 6 |
| | GPLv2+/3+ | No license | 8 | 48 | 7 |
| | LesserGPLv2.1+/3+ | No license | 16 | | 9 |
| | MPLv1.1 | No license | | | 1 |
| | Oracle | No license | | 3 | |
| | Proprietary | No license | | 1 | 2 |
| | Sun Microsystems | No license | | 1 | 2 |
| | Unknown | No license | | 11 | |
| | LesserGPLv2.1+ | New BSD3 | 1 | | |
| Total | | | 86 | 78 | 50 |

> are found in 130,719 GitHub projects without evidence of copying, which of 12 were buggy. A toxic code snippet with a race condition was found in two popular projects: deeplearning4j and Apache Hive.

### 3.5 RQ5: Software Licensing Violations

*Do licensing conflicts occur between Stack Overflow clones and their originals?*

In our study, we reveal another type of toxic code snippets which is software licensing issues caused by code cloning to Stack Overflow. We found evidence that 153 pieces of code have been copied from Qualitas projects to Stack Overflow as examples. Another 109 pieces of code are cloned from external sources. Their status of accepted answers increases their chances of being reused. Even though most of the Qualitas projects came with a software license, we found that the license information was frequently missing after the code was copied to Stack Overflow. The licensing terms on top of source code files are not copied because usually only a small part of the file was cloned. In overall, we can see that most of the Stack Overflow snippets do not contain licensing terms while their clone counterparts in Qualitas projects and external sources do. Since licensing statement resides on top of a file, the results here are analysed at a file level, not clone fragment, and clone pairs from the same file are merged. The summary of licensing information is listed in Table 22.

**Compatible license:** There are 41 pairs which have compatible licenses such as *Apache license v.2*; *Eclipse Public License v.1 (EPLv1)*; or a pair of *Creative Common Attribution-NonCommercial-ShareAlike 3.0 Unported (CC BY-NC-SA 3.0)* vs. no license. These clones are safe for being reused. Since source code and text on Stack Overflow are protected by *CC BY-NC-SA 3.0*, we can treat the Stack Overflow code snippets without licensing information as having *CC BY-NC-SA 3.0* by default. The *CC BY-NC-SA 3.0* license is relaxed, and it only requests an attribution when reused.

---

14. The answers were not marked as accepted so they were not included in our experiment.
15. deeplearning4j bug report: https://github.com/deeplearning4j/deeplearning4j/issues/4692
16. Apache Hive bug report: https://issues.apache.org/jira/browse/HIVE-18929

| No. of stars | Qualitas | | Other Projects | | |
|---|---|---|---|---|---|
| | Projects | Pairs | Projects | Pairs | Same license |
| 29,540–10 | 8 | 71 | 406 | 1,837 | 193 |
| 9–5 | 0 | 0 | 275 | 739 | 110 |
| 4–1 | 2 | 24 | 1,746 | 4,536 | 692 |
| Total | 10 | 95 | 2,427 | 7,112 | 995 |

Table 23: Clones of the 214 Stack Overflow missing-license code snippets in 130,719 GitHub projects

**Incompatible license:** there are 214 clone pairs which do not contain licensing information after they are posted on Stack Overflow or contain a different license from their Qualitas clone counterparts. Almost all (85) of **QS** clone pairs have their licensing terms removed or changed when posted on Stack Overflow. One QS clone pair posted by a JFreeChart developer changed its license from Lesser GPL v2.1+ to New BSD 3-clause. The developer may intentionally changed the license to be more suitable to Stack Overflow since New BSD 3-clause license allows reuse without requiring the same license in the distributing software or statement of changes.

For **EX** clone pairs, we searched for licensing terms of the original source code from the external sources. We found that 78 out of 93 EX clone pairs have incompatible licenses. Similarly, the license statement was removed from Stack Overflow snippets.

Of 53 **UD** clone pairs, 50 pairs have incompatible licenses. Again, most clones in Qualitas contain a license while the Stack Overflow snippets do not.

The same GitHub study has been done for license-incompatible code snippets. We detected clones between the 214 code snippets with their original license missing (86 QS, 78 EX, and 50 UD) and 130,719 GitHub projects using SourcererCC with 80% similarity threshold. Opposite to the outdated clones, we discovered a large number of 7,207 clone pairs. There were 95 pairs from 10 Qualitas projects hosted on GitHub and 7,112 pairs from 2,427 other projects. As shown in Table 23, the clones were found in highly-starred projects (29,465 stars) to lowly-starred star projects (1 star). We found 12 code snippets with attributions to Stack Overflow questions/answers and 6 of them refer to one of our QS or EX clone pairs. We used the Ninka tool to identify software licenses of the 7,112 cloned code snippets automatically. Five code snippets did not have a license while the originals had the Apache-2, GPLv2, or EPLv1 license. One snippet had the AGPLv3 license while the original had the Apache-2 license. Only 995 code snippets in GitHub projects have the same license as the originals in Qualitas.

Note that the code snippets could potentially violate the license, but do not necessarily do so. In the example where the JFreeChart developer copied his own code, she or he was free to change the license. The same may have occurred with any of the 214 code snippets.

> For RQ5, we found 214 code snippets on Stack Overflow that could potentially violate the license of their original software. The majority of them do not contain licensing statements after they have been copied to Stack Overflow.

> For 164 of them, we were able to identify, with evidence, where the code snippet has been copied from. We found occurrences of 7,112 clones of the 214 license-incompatible code snippets in 2,427 GitHub projects.

### 3.6 Overall Discussion

This study empirically shows from the surveys and the clone detection experiment that online code clones occur on Stack Overflow and the clones may become toxic due to outdated code and software license incompatibility. The findings lead to the insight about the toxicity of online code clones, and we proposed three actionable items to the software engineering research and the Stack Overflow community.

#### 3.6.1 Toxicity of outdated and license-incompatible clones

The insights from our study of toxic code snippets on Stack Overflow are as follows:

**Outdated clones are not harmful:** We found only a small number of toxic outdated code snippets in open source projects on GitHub. Besides 12 buggy and outdated code snippets found in 12 projects, the rest were non-harmful clones of the outdated code. Although other studies show that Stack Overflow code snippets may become toxic by containing security vulnerabilities [2], [20] or API misuse [83], we found in this study that the damage caused by outdated code on Stack Overflow is not high.

**License-incompatible clones can be harmful:** The missing licensing statements of online code clones on Stack Overflow can cause more damage than outdated code. As shown in our study and also in the study by An et al. [3], some online clones on Stack Overflow are initially licensed under more restrictive license than Stack Overflow's CC BY-SA 3.0. If these missing-license online clones are reused in software with an incompatible license, the software owner may face legal issues. Software auditing services such as Black Duck Software[17] or nexB[18], which can effectively check for license compliance of code copied from open source projects, do not check for the original license of the cloned code snippets on Stack Overflow. Although the Stack Overflow answerers who participated in our survey believe that most of the code snippets on Stack Overflow are too small to claim for copyright and they fall under fair-use, there is still a risk due to different legal systems in each country. For example, Germany's legal system does not have a concept of fair use. Besides, the number of minimum lines of code to be considered copying, i.e., de minimis, is also differently interpreted from case to case or from country to country.

#### 3.6.2 Actionable Items

Our study discovers links from code in open source projects to code snippets on Stack Overflow using clone detection techniques. These links enable us to discover toxic code snippets with outdated code or licensing problems. The links can be exploited further to mitigate the problems of reusing outdated online clones and incompatible license on

---

17. https://www.blackducksoftware.com
18. https://www.nexb.com

Stack Overflow code snippets. We propose the following actionable items:

**Automated cloning direction detection:** As shown in this study, we still relied on manual validation to read the post, understand the context, and use multiple hints based on human judgement (e.g., comments in the code, natural text in the question and answers, date/time of the posts) to conclude that the code snippets are actually copied from Qualitas or external sources to Stack Overflow. This is a burdensome and time-consuming process. We call for more research in automation of code cloning direction detection. From our experience in this paper, we found that code comments and accompanied natural text on Stack Overflow were a great source of information to decide the direction of code copying. Thus, by using code clone detection to locate clone candidates and then applying information retrieval techniques, e.g., cosine similarity with tf-idf, we can rank the clone candidates based on the similarity of their project names (or classes) to the text in comments or natural text surrounding the clones in Stack Overflow posts. For example, the Stack Overflow answer containing the text *"Actually, you can learn how to compare in* `Hadoop` *from* `WritableComparator`. *Here is an example that borrows some ideas from it."* must be ranked very high among the list of clone candidates of a code snippet from `Hadoop` since it contains two terms of the project name (Hadoop) and a class name (WritableComparator) in it. This technique will dramatically reduce the manual validation effort to establish the direction of cloning. The technique can also be used on Stack Overflow to flag that an answer has a high chance of copying from open source projects.

**Preventive measure:** We encourage Stack Overflow to enforce attribution when source code snippets have been copied *from* licensed software projects to Stack Overflow. Moreover, an IDE plug-in that can automatically detect pasted source code and follow the link to Stack Overflow and then to the original open source projects could also prevent the issue of license violation. We foresee the implementation of the IDE plugin using a combination of scalable code clone detection [64] or clone search techniques [34] and automated software licensing detection [23]. In this study, we performed the check using a set of code clone detectors (Simian and SourcererCC) and software licensing detector (Ninka), but we had to operate the tools manually. Using the knowledge obtained from this study, we plan to build an automated and scalable clone search with licensing detection. With the proposed solution, we can use the tool to create a database of code snippets on Stack Overflow and allow the users to search for clones and check their licenses. The clone search tool can offer a service via REST API and integrated into the IDE plugin. Every time a new code fragment is pasted into the IDE, the plugin performs the check by calling the clone search tool service and report the finding to the developers in real time.

Also, we also performed a study of two open source software auditing platforms/services: BlackDuck Software and nexB. For BlackDuck Software, we found from their report [17] that while they check for code copied from open source projects including GitHub and Stack Overflow and analyse their license compatibility with their customer software, the BlackDuck auditing system will treat the code snippets on Stack Overflow as having an "unknown" license because it does not know the original license of Stack Overflow code snippets. For nexB, their product does not mention checking of reused source code from Stack Overflow. So, our proposed service, which can offer more precise licensing information of Stack Overflow code snippets, will be useful as an add-on license check for code copying from Stack Overflow.

**Detective measure:** A system to detect outdated source code snippets on Stack Overflow may be needed. The system can leverage the online clone detection techniques in this study to periodically check if the cloned snippets are still up-to-date with their originals.

While checking if the copied code snippets on Stack Overflow are still up-to-date with the latest version of their originals can possibly be done automatically, it is a challenging task to automate the process of establishing the direction of code cloning as previously discussed. Since automatically establishing code cloning direction is still an open challenge, one viable solution, for now, is encouraging the Stack Overflow developers to always include the origin of the copied code snippet in the post so that this link is always established at the posting time. Even better, Stack Overflow can provide an optional form to fill in when an answerer post an answer if he or she copies the code from other software projects. The form should include the origin of the code snippet (possibly as a GitHub URL) and its original license. Using this manually established links at posting time, we can then automate the process of checking for an outdated code.

With such a system, the poster can be notified when the code has been updated in the original project so that he/she can update their code on Stack Overflow accordingly. On the other hand, with a crowdsourcing solution using an IDE plug-in, developers can also report the corrected version of outdated code back to the original Stack Overflow threads when they reuse outdated code and make corrections to them.

## 4 THREATS TO VALIDITY

**Internal validity:**

We applied different mechanisms to ensure the validity of the clone pairs we classified. First, we used two widely-used clone detection tools, Simian and SourcererCC. We tried five other clone detectors but could not add them to the study due to their scalability issues and susceptibility to incomplete code snippets. We adopted Bellon's agreement metric [8] to merge clone pairs for the manual classification and avoid double counting of the same clone pairs. We studied the impact of choosing different thresholds for Bellon's clone agreement and the minimum clone size of the two clone detectors and selected the optimal values. Nevertheless, our study might still suffer from false negatives, i.e., online code clones that are not reported by the tools or are filtered out by the size (less than 10 lines) within the clone detection process. We selected accepted answers on Stack Overflow in this study to focus on code snippets that solve the question's problem and are often shown on top of the answer list. We investigated the 72,365 Stack Overflow code snippets used in our study and found that 62,245 of them (86%) are also the highest voted answers. We plan to incorporate the highest voted answers in future work. We analysed the Stack Overflow data snapshot of January 2016. The findings may differ from using the newer snapshots. However, we do not expect large differences of the findings if we update the data set. The reason is that the Stack Overflow answers are rarely modified. Moreover, updating the dataset would eliminate the possibility to study the evolution of outdated answers.

Our seven patterns of online code cloning may not cover all possible online cloning patterns. However, instead of defining the patterns beforehand, we resorted to extracting them from the data sets. We derived them from a manual investigation of 679 online clone pairs and adopted one pattern from the study by Kapser et al. [31].

The 2,289 clone pairs classified by the first and the third author are subject to manual judgement and human errors. Although we tried our best to be careful on searching for evidence and classifying the clones, some errors may still exist. We mitigated this problem by having two investigators to cross check the classifications and found 145 cases that lead to better classification results. This validation process can be even improved by employing an external investigator.

The detailed investigation of the 100 outdated answers on Stack Overflow may not fully capture what Stack Overflow visitors actually do when they already expect an outdated answer. Our investigation tries to at least shade light on this by comparing the popularity of the outdated answer and the newer answers based on the number of votes. Nonetheless, this may be affected by the duration that the answers appear on a Stack Overflow post. The newer answers have less visibility compared to the older answers.

**External validity:** We carefully chose the data sets for our experiment so the findings could be generalised as much as possible. We selected Stack Overflow because it is one of the most popular programming Q&A websites available with approximately 7.6 million users. There are a large number of code snippets reused from the site [3], and there are also several studies encouraging of doing so (e.g., [33], [49], [52], [53]). Nonetheless, it may not be representative to all the programming Q&A websites.

Regarding the code snippets, we downloaded a full data dump and extracted Java accepted answers since they are the most likely ones to be reused. Our findings are limited to these restrictions. They may not be generalised to all programming languages and all answers on Stack Overflow. We chose the curated Qualitas corpus for Java open source projects containing 111 projects [73]. The projects span several areas of software and have been used in several empirical studies [7], [48], [72], [76]. Although it is a curated and well-established corpus, it may not fully represent all Java open source software available.

We selected 130,719 GitHub Java projects based on the number of stars they obtained to represent their popularity. They might not represent all Java projects on GitHub, and the number of clone pairs found may differ from other project selection criteria.

Regarding the few number of toxic code snippets in Stack Overflow posts and code reuse in GitHub, this is partially due to the data set of (only) 111 Java projects from the Qualitas corpus. The clones and the discovered outdated code and potentially license-violating code snippets reported in this paper are only subject to these 111 projects. The number may increase if we expand the amount of open source projects. Moreover, this study only focuses on Java, and the situation may be worse or better for other programming languages.

## 5 RELATED WORK

**Stack Overflow** is a gold mine for software engineering research and has been put to use in several previous studies. Regarding developer-assisting tools, Seahawk is an Eclipse plug-in that searches and recommends relevant code snippets from Stack Overflow [52]. A follow-up work, Prompter, by Ponzanelli et al. [53] achieves the same goal but with improved algorithms. The code snippets on Stack Overflow are mostly examples or solutions to programming problems. Hence, several code search systems use whole or partial data from Stack Overflow as their code search databases [16], [33], [49], [69], [70]. Furthermore, Treude et al. [74] use machine learning techniques to extract insight sentences from Stack Overflow and use them to improve API documentation.

Another research area is knowledge extraction from Stack Overflow. Nasehi et al. [47] studied what makes a good code example by analysing answers from Stack Overflow. Similarly, Yang et al. [81] report the number of reusable code snippets on Stack Overflow across various programming languages. Wang et al. [77] use Latent Dirichlet Allocation (LDA) topic modelling to analyse questions and answers from Stack Overflow so that they can automatically categorise new questions. There are also studies trying to understand developers' behaviours on Stack Overflow, e.g., [9], [12], [46], [61].

**Code clone detection** is a long-standing research topic in software engineering. Whether clones are good or bad for software is still controversial [24], [26], [27], [31], [32], [40], [64]. Code clones have several applications such as software plagiarism detection [54], source code provenance [14], and software licensing conflicts [22].

Two code fragments are clones if they are similar enough according to a given definition of similarity [8]. Given an open interpretation of "definition of similarity," there are various clone detection tools and their siblings, code plagiarism detectors, invented based on a plethora of different code similarity measurements [57], [58], [62], [71]. Many tools do not work on original source code directly but detect clones at an intermediate representation such as tokens [10], [18], [25], [29], [44], [54], [64], [65], [67], AST [6], [28] or program dependence graphs [36], [38].

**Cloning patterns** are initially defined by Kapser et al. [31], [32] by studying clones in Linux file systems and deriving 11 patterns of code cloning. Our study adopted one of the patterns into our online code cloning patterns.

**Clone agreement** is useful when a clone oracle is absent. By exploiting the different behaviours of clone detectors, one can look for their agreement and obtain highly-confident clones [8], [79]. Clone pairs that are agreed by multiple tools are more assured to be true clones than the ones reported by only a single tool [21], [60], [79].

**Software licensing** is crucial for open source and industrial software development. Di Penta et al. [15] studied the evolution of software licensing in open source software and found that licensing statements change over time. German et al. [22] found that licensing conflicts occur between the clone siblings, i.e., clones among different systems that come from the same source. Later, German et al. [23] created an automated tool for software license identification, Ninka, which is used in our online clone license analysis.

**Reusing of outdated third-party source code** occurs in software development. Xia et al. [80] show that a large number of open source systems reuse outdated third-party libraries from popular open source projects. Using the outdated code give detrimental effects to the software since they may introduce vulnerabilities. Our study similarly finds the outdated code on Stack Overflow.

Work similar to ours are studies by An et al. [3], Abdalkareem et al. [1], Baltes et al. [4], and Zhang et al. [83]. An et al. investigated clones between 399 Android apps and Stack Overflow posts. They found 1,226 code snippets which were reused from 68 Android apps. They also observed that there are 1,279 cases of potential license violations. The authors rely on the timestamp to judge whether the code has been copied from/to Stack Overflow along with confirmations from six developers. Instead of Android apps, we investigated clones between Stack Overflow and 111 open source projects. Their results are similar to our findings that there exist clones from software projects to Stack Overflow with potential licensing violations. Abdalkareem et al. [1] detected clones between Stack Overflow posts and Android apps from the F-Droid repository and used timestamps to determine the direction of copying. They found 22 apps containing code cloned from Stack Overflow. They reported that cloned code is commonly used for enhancing existing code. Their analysis shows that the cloned code from Stack Overflow has detrimental effects on quality of the apps. The median percentage of bug fixing commits after adding Stack Overflow code (19.09%) is higher than before adding the code (5.96%) with statistical significance. Baltes et al. [4] discovered that only 23.2% of the clones in GitHub projects from the 10 most frequently referenced Java code snippets on Stack Overflow contain attributions. Zhang et al. [83] study quality of code snippets on Stack Overflow. They show that 31% of the analysed Stack Overflow posts contain potential API usage violations and could lead to program crashes or resource leaks.

## 6 CONCLUSIONS

Online code clones are clones that have been copied to Q&A websites such as Stack Overflow. We classified 2,289 clone pairs using seven patterns of online code cloning. We discovered 153 clone pairs that have been copied, with evidence, from Qualitas projects to Stack Overflow, 109 clone pairs that have been copied from external sources besides Qualitas to Stack Overflow, and 65 clone pairs that are highly similar but without evidence of copying.

The online survey of 201 high-reputation Stack Overflow developers (i.e., answerers) shows that although the developers are aware of outdated code in their answers, 19.8% of them rarely or never fix the outdated code. In addition, 62% of the participants are aware of Stack Overflow CC BY-SA 3.0 license applied to code snippets on the website. Only 3 answerers include the original license in their answers. 69% of the answerers never check for licensing conflicts between the original code and CC BY-SA 3.0 enforced by Stack Overflow. Another survey of 87 Stack Overflow visitors shows that Stack Overflow code snippets have several issues, including outdated code. 85% of them are not aware of CC BY-SA 3.0 license enforced by Stack Overflow and 66% never check for license conflicts when reusing code snippets.

We support the surveys' findings by performing a detailed analysis of toxic code snippets on Stack Overflow. We investigated the online clone pairs on two aspects: outdated code and potential license violation. The investigation of the 153 clone pairs copied, with evidence, from Qualitas to Stack Overflow reveals that 100 of them are outdated. Twelve outdated clone pairs are buggy and toxic to reuse. Our large-scale clone detection between the outdated code snippets and 130,719 GitHub projects finds 102 candidates of the outdated code being used in the wild. Moreover, we found 214 code snippets on Stack Overflow that could potentially violate the license of their original software, and they occur in 7,112 times in 2,427 projects on GitHub.

This study is among, if not the first, to address the important issues of toxic code snippets, including outdated and license-violating online code clones, on programming Q&A websites using a hybrid methodology of automatic code clone detection and a manual clone investigation.

## ACKNOWLEDGMENTS

The authors would like to thank Prof. Cristina Lopes and Di Yang from University of California, Irvine for their help in running SourcererCC clone detector and implementing a custom tokeniser for Stack Overflow snippets.

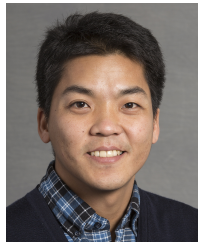

**Chaiyong Ragkhitwetsagul** is a lecturer at the Faculty of Information and Communication Technology, Mahidol University, Thailand. He received the PhD degree in Computer Science at University College London, where he was part of the Centre for Research on Evolution, Search, and Testing (CREST). His research interests include code search, code clone detection, software plagiarism, modern code review, and mining software repositories.

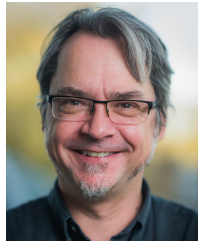

**Jens Krinke** is Associate Professor in the Software Systems Engineering Group at the University College London, where he is Director of CREST, the Centre for Research on Evolution, Search, and Testing. His main focus is software analysis for software engineering purposes. His current research interests include software similarity, modern code review, program analysis, and software testing. He is well known for his work on program slicing and clone detection.

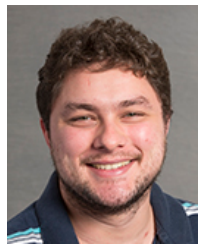

**Matheus Paixao** is Research Assistant in the Computer Science Department at State University of Ceara. He received his PhD degree at University College London, where he was part of the Centre for Research on Evolution, Search, and Testing (CREST) and Software Systems Engineering (SSE) Group. His research interests include software architecture, search-based software engineering, mining software repositories, and modern code review.

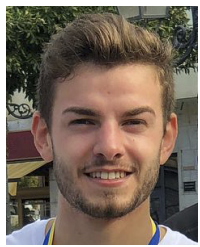

**Giuseppe Bianco** received his bachelor degree in Computer Science from the University of Molise (Italy) under the supervision of Dr. Rocco Oliveto. As part of an Erasmus+ traineeship, he spent three month at the CREST centre, University College London, UK, under the supervision of Dr. Jens Krinke. He is now a project manager at Corsi.it.

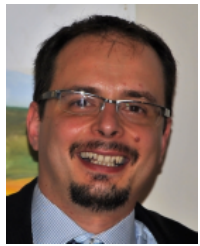

**Rocco Oliveto** is Associate Professor at University of Molise (Italy), where he is also the Director of the Computer Science Bachelor and Master programs and the Director of the Software and Knowledge Engineering Lab (STAKE Lab). He is also one of the co-founders and CEO of DataSound, a spin-off of the University of Molise aiming at efficiently exploiting the priceless heritage that can be extracted from big data analysis via machine learning.